# Photon Wave Mechanics

Zhi-Yong Wang[1*], Cai-Dong Xiong[1], Ole Keller[2]

[1]School of Optoelectronic Information, University of Electronic Science and Technology of China, Chengdu 610054, CHINA

[2]Institute of Physics, Aalborg University, Pontoppidanstrcede 103, DK-9220 AalborgØst, Denmark

E-Mail: zywang@uestc.edu.cn

In contrast to wave functions in nonrelativistic quantum mechanics interpreted as probability amplitudes, wave functions in relativistic quantum mechanics have generalized meanings such as charge-density amplitudes, energy-density amplitudes as well as particle-number density amplitudes, etc. Applying electromagnetic field intensities we construct a photon wave function, it corresponds to the (1,0)+(0,1) spinor representation of the electromagnetic field, and can be interpreted as the energy-density amplitude of photons outside a source. In terms of photon wave functions we develop photon wave mechanics, which provides us with a new quantum-mechanical description for photons outside a source.

**PACS:**  42.50.Ct; 03.50.De; 03.65.Ca

In a recent comprehensive report [1] one of the authors (Ole Keller) makes a thorough study of the spatial localization of photons, where the traditional photon wave mechanics developed in the previous literatures [2-8], is systematically reviewed. In fact, in spite of QED's great success as well as the traditional conclusion that single photon cannot be localized [9-10], there have been many attempts to develop photon wave mechanics which is based on the concept of photon wave function and related to the first quantization theory of the photon [2-8]. These efforts have both theoretical and practical interests. For example, contrary to the traditional opinion, some recent experimental and theoretical studies have shown that photons can be localized in space [11-13].

However, the traditional photon wave mechanics has some shortcomings: 1). the negative-energy solution is directly dismissed (but for the case that, let the negative- and positive-energy solutions describe the left- and right-handed circular polarization components, respectively), this is problematic. For example, without the negative-energy part, one cannot carry through the canonical quantization of the photon field, because all the equal-time commutators for the field operators and their canonically conjugate momenta vanish. In the absence of the negative frequency solution, the Hamiltonian does not involve zero-point energy, which conflicts with the QED; 2). In the previous literatures, the solution describing the admixture of the longitudinal and scalar photon states is usually discarded artificially, such that the completeness is not met; 3). The traditional photon wave mechanics is short of a standard theoretical framework, and photon wave functions are obtained usually by a tentative constructing, rather than by solving the corresponding quantum-mechanical equation in a rigorous and unified way.

In view of mentioned above, in this letter, we will try to provide a systemically formulation for photon wave mechanics, at both levels (first and second) of quantization. In our framework, the negative-energy solution is reasonably preserved, which is in agreement with the standard relativistic quantum theory; the solution describing the admixture of the longitudinal and scalar photon states is kept and physically its contribution to energy and momentum is discarded in a natural way. Furthermore, in our case, photon wave function is taken as a 6×1 spinor that transforms according to the $(1,0) \oplus (0,1)$ representation of the Lorentz group. In the following we apply the natural units of measurement ($\hbar = c = 1$) and the space-time metric tensor is taken as $g_{\mu\nu} = \text{diag}(1,-1,-1,-1)$. Repeated indices must be summed according to the Einstein rule. In this letter we let $x^\mu = (t,-\boldsymbol{x})$, instead of $x^\mu = (t,\boldsymbol{x})$, denote the contravariant position 4-vector (and so on).

In vacuum the electric field, $\boldsymbol{E} = (E_1, E_2, E_3)$, and the magnetic field, $\boldsymbol{B} = (B_1, B_2, B_3)$,

satisfy the Maxwell equations

$$\nabla \times \boldsymbol{E} = -\partial_t \boldsymbol{B}, \quad \nabla \times \boldsymbol{B} = \partial_t \boldsymbol{E}, \tag{1}$$

$$\nabla \cdot \boldsymbol{E} = 0, \quad \nabla \cdot \boldsymbol{B} = 0. \tag{2}$$

Let $(\tau_i)_{jk} = -i\varepsilon_{ijk}$, $i, j, k = 1, 2, 3$, where $\varepsilon_{ijk}$ is a totally antisymmetric tensor with $\varepsilon_{123} = 1$, by them we define matrix vector $\boldsymbol{\tau} = (\tau_1, \tau_2, \tau_3)$, where

$$\tau_1 = \begin{pmatrix} 0 & 0 & 0 \\ 0 & 0 & -i \\ 0 & i & 0 \end{pmatrix}, \tau_2 = \begin{pmatrix} 0 & 0 & i \\ 0 & 0 & 0 \\ -i & 0 & 0 \end{pmatrix}, \tau_3 = \begin{pmatrix} 0 & -i & 0 \\ i & 0 & 0 \\ 0 & 0 & 0 \end{pmatrix}.$$

By means of matrix vector $\boldsymbol{\tau} = (\tau_1, \tau_2, \tau_3)$ and the quantities ($I_{n \times n}$ denotes the $n \times n$ unit matrix, $n = 2, 3, 4...$, T denotes the transpose, $\chi = \beta_0 \boldsymbol{\beta}$)

$$\beta_0 = \begin{pmatrix} I_{3\times 3} & 0 \\ 0 & -I_{3\times 3} \end{pmatrix}, \quad \boldsymbol{\beta} = \begin{pmatrix} 0 & \boldsymbol{\tau} \\ -\boldsymbol{\tau} & 0 \end{pmatrix}, \tag{3}$$

$$E = \begin{pmatrix} E_1 & E_2 & E_3 \end{pmatrix}^T, \quad B = \begin{pmatrix} B_1 & B_2 & B_3 \end{pmatrix}^T, \tag{4}$$

$$\psi = \frac{1}{\sqrt{2}} \begin{pmatrix} E \\ iB \end{pmatrix}, \tag{5}$$

one can rewrite the Maxwell equations as a *Dirac-like equation*

$$i\beta^\mu \partial_\mu \psi(x) = 0, \text{ or } i\partial_t \psi(x) = \hat{H}\psi(x), \tag{6}$$

where $\hat{H} = -i\boldsymbol{\chi} \cdot \nabla$ represents the Hamiltonian of the free photon. One can show that the $6 \times 1$ spinor $\psi(x)$ transforms according to the $(1,0) \oplus (0,1)$ representation of the Lorentz group. Using Eq. (2) one can show that $(\beta^\mu \partial_\mu)(\beta_\nu \partial^\nu) = \partial^\mu \partial_\mu + \Omega$ with $\Omega\psi(x) = 0$, thus Eq. (6) implies that the wave equation $\partial^\mu \partial_\mu \psi(x) = 0$. Note that the equation $\Omega\psi(x) = 0$ corresponds to the transverse conditions given by Eq. (2), where

$$\Omega = I_{2\times 2} \otimes [\begin{pmatrix} \partial_1 \\ \partial_2 \\ \partial_3 \end{pmatrix} \begin{pmatrix} \partial_1 & \partial_2 & \partial_3 \end{pmatrix}], \tag{7}$$

where $\otimes$ denotes the direct product, and $\partial_i = \partial/\partial x^i$, $i = 1, 2, 3$. Let $\hat{\boldsymbol{L}} = \boldsymbol{x} \times (-i\nabla)$ be the orbital angular momentum operator, one can easily obtain $[\hat{H}, \hat{\boldsymbol{L}} + \boldsymbol{S}] = 0$, where $\boldsymbol{S} = I_{2\times 2} \otimes \boldsymbol{\tau}$ satisfying $\boldsymbol{S} \cdot \boldsymbol{S} = 1(1+1)I_{6\times 6}$ represents the spin matrix of the spin-1 field.

Let $k_\mu = (\omega, \boldsymbol{k})$ denote the 4-momentum of photons ($\hbar = c = 1$), where $\omega$ is the frequency and $\boldsymbol{k}$ the wave-number vector. The fundamental solutions of the Dirac-like equation (6) are represented by the positive- and negative-frequency components, respectively

$$\begin{cases} \phi_{k,\lambda}(x) = (\omega/V)^{1/2} f(k,\lambda) \exp(-ik \cdot x) \\ \phi_{-k,\lambda}(x) = (\omega/V)^{1/2} g(k,\lambda) \exp(ik \cdot x) \end{cases}, \qquad (8)$$

where $\int (1/V) \mathrm{d}^3 x = 1$, $\lambda = \pm 1, 0$, and

$$f(k,\lambda) = \frac{1}{\sqrt{1+\lambda^2}} \begin{pmatrix} \varepsilon(\boldsymbol{k},\lambda) \\ \lambda \varepsilon(\boldsymbol{k},\lambda) \end{pmatrix}, \qquad (9)$$

$$g(k,\lambda) = \frac{1}{\sqrt{1+\lambda^2}} \begin{pmatrix} \lambda \varepsilon(\boldsymbol{k},\lambda) \\ \varepsilon(\boldsymbol{k},\lambda) \end{pmatrix}. \qquad (10)$$

In matrix form, $\varepsilon(\boldsymbol{k}, 0) = (k_1 \ k_2 \ k_3)^T / |\boldsymbol{k}|$ is the longitudinal polarization vector of photons, while $\varepsilon(\boldsymbol{k}, 1)$ and $\varepsilon(\boldsymbol{k}, -1) = \varepsilon^*(\boldsymbol{k}, 1)$ (the complex conjugate of $\varepsilon(\boldsymbol{k}, 1)$) are the right- and left-hand circular polarization vectors, respectively, where

$$\varepsilon(\boldsymbol{k}, 1) = \varepsilon^*(\boldsymbol{k}, -1) = \frac{1}{\sqrt{2}|\boldsymbol{k}|} \begin{pmatrix} \dfrac{k_1 k_3 - i k_2 |\boldsymbol{k}|}{k_1 - i k_2} \\ \dfrac{k_2 k_3 + i k_1 |\boldsymbol{k}|}{k_1 - i k_2} \\ -(k_1 + i k_2) \end{pmatrix}. \qquad (11)$$

Correspondingly, $\lambda = \pm 1, 0$ represent the spin projections in the direction of $\boldsymbol{k}$. In fact, when electromagnetic fields are described by the electromagnetic 4-potential, there involves four polarization vectors describing four kinds of photons, while described by the photon wave function (constructed by the electromagnetic field intensities), there only involves three polarization vectors. Therefore, in our framework, the $\lambda=0$ solution describes the admixture of the longitudinal and scalar photon states, while the $\lambda=\pm 1$ solutions describe the transverse photon states.

Let $\varepsilon^+$ stand for the Hermitian conjugate of $\varepsilon$ (and so on), one has

$$\begin{cases} \varepsilon^+(\boldsymbol{k},\lambda)\varepsilon(\boldsymbol{k},\lambda')=\delta_{\lambda\lambda'} \\ \sum_\lambda \varepsilon(\boldsymbol{k},\lambda)\varepsilon^+(\boldsymbol{k},\lambda)=I_{3\times 3} \end{cases}. \qquad (12)$$

The fundamental solutions given by Eq. (8) satisfy the orthonormality and completeness relations

$$\begin{cases} \int \phi_{k,\lambda}^+ \phi_{k',\lambda'} \mathrm{d}^3 x = \int \phi_{-k,\lambda}^+ \phi_{-k',\lambda'} \mathrm{d}^3 x = \omega \delta_{\lambda\lambda'} \delta_{\boldsymbol{k}\boldsymbol{k}'} \\ \int \phi_{k,\lambda}^+ \phi_{-k_0,k',\lambda'} \mathrm{d}^3 x = \int \phi_{-k_0,k,\lambda}^+ \phi_{k',\lambda'} \mathrm{d}^3 x = 0 \end{cases}, \qquad (13)$$

$$\sum_\lambda \int (\phi_{k,\lambda} \phi_{k,\lambda}^+ + \phi_{-k_0,k,\lambda} \phi_{-k_0,k,\lambda}^+) \mathrm{d}^3 x = \omega I_{6\times 6}. \qquad (14)$$

Substituting $\varphi(k)\exp(-\mathrm{i}k^\mu x_\mu)$ into Eq. (6) one has $\det(\omega-\boldsymbol{\chi}\cdot\boldsymbol{k})=0$. Let $k_\mathrm{T}^\mu=(\omega_\mathrm{T},-\boldsymbol{k}_\mathrm{T})$ denote the 4-momenta of the transverse photons (corresponding to the $\lambda=\pm 1$ solutions), while $k_\mathrm{L}^\mu=(\omega_\mathrm{L},-\boldsymbol{k}_\mathrm{L})$ the 4-momenta of the longitudinal and scalar photons (both corresponding to the $\lambda=0$ solution), using $\det(\omega-\boldsymbol{\chi}\cdot\boldsymbol{k})=0$ and Eq. (2) one has

$$\omega_{\pm 1}\equiv\omega_\mathrm{T}=|\boldsymbol{k}_\mathrm{T}|, \quad \omega_0\equiv\omega_\mathrm{L}=|\boldsymbol{k}_\mathrm{L}|=0, \qquad (15)$$

which in agreement with the traditional conclusions that the contributions of the longitudinal and scalar photons to the energy and momentum cancel each other. For convenience the $\lambda=0$ solution is kept and only in the end let $k_L^\mu \to 0$ for the $\lambda=0$ solution.

Consider that antiphotons are identical with photons, we expand the general solution of Eq. (6) via the fundamental solutions $\phi_{k,\lambda}$ and $\phi_{-k,\lambda}$ as

$$\psi(x) = \frac{1}{\sqrt{2}} \sum_{k,\lambda} [a(k,\lambda)\phi_{k,\lambda} + a^{+}(k,\lambda)\phi_{-k,\lambda}], \tag{16}$$

it corresponds to the $(1,0) \oplus (0,1)$ representation of the Lorentz group, such that one can prove that

$$\mathcal{L} = \bar{\psi}(x)(i\beta^{\mu}\partial_{\mu})\psi(x) \tag{17}$$

is a Lorentz scalar (see **Appendix A**), but its dimension is $[1/\text{length}]^5$ and then we call it pseudo-Lagrangian density. Applying the variational principle in $A = \int \mathcal{L} d^4 x$ one can obtain the Dirac-like equation (6). In the present case, in terms of the inverse of the operator $-i\partial_t$, i.e., $(-i\partial_t)^{-1}$, we define the canonical momentum conjugating to $\psi(x)$ as ($\dot{\psi} = \partial_t \psi$)

$$\pi \equiv (-i\frac{\partial}{\partial t})^{-1} \frac{\partial \mathcal{L}}{\partial \dot{\psi}} = \frac{\partial \mathcal{L}'}{\partial \dot{\psi}} = (-i\partial_t)^{-1} i\psi^{+}, \tag{18}$$

where

$$\mathcal{L}' \equiv [(-i\partial_t)^{-1}\bar{\psi}(x)]i\beta^{\mu}\partial_{\mu}\psi(x) \tag{19}$$

is called equivalent pseudo-Lagrangian density. The conserved charges related to the invariance of $\mathcal{L}'$ under the space-time translations, are the 4-momentum (say, $p^{\mu} = (H, \boldsymbol{p})$) of the photon field

$$H = \int [\pi(x)\dot{\psi}(x) - \bar{\mathcal{L}}']d^3 x, \tag{20}$$

$$\boldsymbol{p} = -\int [\pi(x)\nabla\psi(x)]d^3 x. \tag{21}$$

In field quantization the electromagnetic 4-potential (say, $A^{\mu}(x)$, $\mu = 0,1,2,3$) satisfies the canonical commutation relations

$$[A^{\mu}(x), A^{\nu}(y)] = -ig^{\mu\nu}D(x-y), \tag{22}$$

where

$$iD(x) \equiv \int \frac{d^3k}{(2\pi)^3} \frac{1}{2\omega}[\exp(-ik \cdot x) - \exp(ik \cdot x)]. \tag{23}$$

Using Eqs. (4)-(5), (22)-(23) and ($i, j, k = 1, 2, 3$)

$$E^i = \partial^i A^0 - \partial^0 A^i, \quad B^i = \varepsilon^{ijk} \partial_j A_k, \tag{24}$$

one can obtain the canonical commutation relations

$$\begin{cases} [\psi_i(\mathbf{x},t), \pi_j(\mathbf{x}',t)] = -i\delta_{Tij}\delta^3(\mathbf{x}-\mathbf{x}')/2 \\ [\psi_{i+3}(\mathbf{x},t), \pi_{j+3}(\mathbf{x}',t)] = -i\delta_{Tij}\delta^3(\mathbf{x}-\mathbf{x}')/2 \end{cases}, \tag{25}$$

the other commutators vanish, where $i, j = 1, 2, 3$ and $\delta_{Tij} \equiv \delta_{ij} - (\partial_i \partial_j / \nabla^2)$ is the transverse delta function. Using Eqs. (16), (18) and (25), we get the following commutation relations

$$[a(k,\lambda), a^+(k',\lambda')] = \delta_{kk'}\delta_{\lambda\lambda'}, \tag{26}$$

with other commutators vanishing. For the moment, Eq. (6) can be obtained from Heisenberg's equation of motion $\partial_t \psi = i[H, \psi]$.

Using $k_L^\mu = (\omega_L, -\mathbf{k}_L) = 0$ and Eqs. (20), (21), (26), one has

$$H = \sum_k \sum_{\lambda=\pm 1} \omega[a^+(k,\lambda)a(k,\lambda) + (1/2)], \tag{27}$$

$$\mathbf{p} = \sum_k \sum_{\lambda=\pm 1} \mathbf{k}[a^+(k,\lambda)a(k,\lambda)], \tag{28}$$

this is in agreement with the traditional theory. By the way, applying $k_L^\mu = (\omega_L, -\mathbf{k}_L) = 0$ one can prove that the Green function of Eq. (6) is

$$iR_f(x_1 - x_2) = i\beta^\mu \partial_\mu \delta_T i\Delta(x_1 - x_2), \tag{29}$$

where $\delta_T$ is the transverse delta function with components $\delta_{Tij} \equiv \delta_{ij} - (\partial_i \partial_j / \nabla^2)$, $i, j = 1, 2, 3$, and $i\Delta(x)$ is the free propagator of massless scalar fields ($\varepsilon \to 0$):

$$i\Delta(x) = \int \frac{d^4k}{(2\pi)^4} \frac{i}{k^2 + i\varepsilon} \exp(-ik \cdot x). \tag{30}$$

Consider that Eqs. (8)-(10) and (5), one can show that the interchange of the positive- and negative-energy solutions is equivalent to the duality transformation between the electric field $E$ and magnetic field $B$, and then the symmetry between the positive- and negative- energy solutions corresponds to the duality between the electric and magnetic fields, rather than to the usual particle-antiparticle symmetry (see **Appendix B**). Moreover, in the positive-energy solutions $E$ has longitudinal component while $B$ not; in the negative-energy solutions $B$ has longitudinal component while $E$ not. Therefore, if the positive-energy solutions are regarded as the fields produced by electric multipole moments [14], then the negative-energy solutions can be regarded as the fields produced by magnetic multipole moments. In fact, the Dirac-like equation (6) is valid for all kinds of electromagnetic fields outside a source, including the time-varying and static fields, the transverse and longitudinal fields, and those generated by an electrical or magnetic multipole moments, etc. As a proper theory of photon wave mechanics, it naturally contains the negative-energy solutions and the $\lambda = 0$ solution without presenting any problem.

Then, we develop a rigorous theory for photon wave mechanics without resorting to any additional hypothesis, in this theory some defects presenting in previous theory [2-8] are removed and some new insights into the traditional theory are obtained. For example, we give a more natural treatment for the longitudinal and scalar photons, and provide a new perspective on the symmetry between the positive- and negative-energy solutions by associating it with the duality between the electric and magnetic fields. Field quantization is obtained according to our framework. These efforts have potential interests in near-field optics and optical tunneling theory. For example, traditionally, the propagation of guided evanescent waves (i.e., photonic tunneling) is studied just by a quantum-mechanical analogy, while, by photon wave mechanics developed here, one can study it via photon's quantum theory itself. Moreover, to know a single-photon state means to know its electric and magnetic field distributions in space and time, then such a state can be described by the photon wave function introduced here, and its generation is an important goal in quantum-information research [15].

It is important to note that, there is an alternative way of developing our theory: to define an inner product and the mean value of an operator $\hat{L}$ as $\langle\psi|\psi'\rangle \equiv \int \psi^+ \psi' \mathrm{d}^3 x$ and $\langle L \rangle \equiv \int \psi^+ \hat{L} \psi \mathrm{d}^3 x$, respectively, the solutions of the Dirac-like equation (6) can be rechosen as those having the dimension of $[1/\text{length}]^{3/2}$ (rather than $[1/\text{length}]^2$), such that they no longer correspond to the $(1,0) \oplus (0,1)$ representation of the Lorentz group, but rather to particle-number density amplitude. This presents no problem provided that one can ultimately obtain Lorentz-covariant observables. In fact, in nonrelativistic quantum mechanics, the wave functions of all kinds of particles stand for probability amplitudes with the dimension of $[1/\text{length}]^{3/2}$, and do not correspond to any representation of the Lorentz group. In the first-quantized sense, the field quantities presented in our formalism can be regarded as charge-density amplitude or particle-number density amplitude, which is an extension for the concept of probability amplitude. For the moment, the normalization factor presented in $\psi(x)$ is chosen as $1/\sqrt{V}$ instead of the original $\sqrt{\omega}/\sqrt{V}$ (see Eq. (8)), and the factor of $\omega$ no longer appears on the right of Eq. (13) and (14). In particular, now the Lagrangian density $\mathcal{L} = \bar{\psi}(x)(\mathrm{i}\beta^\mu \partial_\mu)\psi(x)$ has the dimension of $[1/\text{length}]^4$, the canonical momentum conjugating to $\psi(x)$ is defined according to the traditional form of $\pi = \partial \mathcal{L}/\partial \dot{\psi} = i\psi^+$, and the free Feynman propagator of the photon is defined as $iR_f(x_1 - x_2) \equiv \langle 0|T\psi(x_1)\bar{\psi}(x_2)|0\rangle$ (it is the Green function of Eq. (6)). However, all the final physical conclusions are preserved.

**Appendix A**

**Lorentz invariance of the pseudo-Lagrangian density**

In our formalism, the photon field as the 6×1 spinor, transforms according to the $(1,0) \oplus (0,1)$ representation of the Lorentz group. As a result, we can demonstrate the Lorentz invariance of the pseudo-Lagrangian density given by Eq. (17). Under the infinitesimal Lorentz transformation

$$x^\mu \to x'^\mu = a^{\mu\nu} x_\nu, \quad \partial_\mu \to \partial'_\mu = a_{\mu\nu} \partial^\nu, \qquad (a\text{-}1)$$

where $a_{\mu\nu} = g_{\mu\nu} + \varepsilon_{\mu\nu}$ with $\varepsilon_{\mu\nu}$ being the infinitesimal antisymmetric tensor, $\psi(x)$ transforms linearly in the way

$$\psi(x) \to \psi'(x') = \Lambda \psi(x), \quad \bar\psi(x) \to \bar\psi'(x') = \bar\psi(x) \beta^0 \Lambda^+ \beta^0, \qquad (a\text{-}2)$$

where $\Lambda = 1 - \frac{i}{2} \varepsilon^{\mu\nu} \Sigma_{\mu\nu}$ with $\Sigma_{\mu\nu}$ being the infinitesimal generators of Lorentz group (in the $(1,0) \oplus (0,1)$ representation). Let $\varepsilon_{lmn}$ denote the full antisymmetric tensor with $\varepsilon_{123} = 1$, one has $\Sigma_{lm} = \varepsilon_{lmn} S_n$ and $\Sigma_{l0} = -\Sigma_{0l} = i\chi_l$ ($l,m,n = 1,2,3$), where $\boldsymbol{S} = I_{2\times 2} \otimes \boldsymbol{\tau}$ $(= (S_1, S_2, S_3))$ is the spin matrix of the photon field, while $\boldsymbol{\chi} = \beta_0 \boldsymbol{\beta}$ plays the role of the infinitesimal generators of Lorentz boost. Under the transformations (a-1) and (a-2), the pseudo-Lagrangian density given by Eq. (17) transforms as

$$\mathcal{L} \to \mathcal{L}_T = \bar\psi(x) \beta^0 \Lambda^+ \beta^0 (i\beta^\mu a_{\mu\nu} \partial^\nu) \Lambda \psi(x), \qquad (a\text{-}3)$$

let $\delta\mathcal{L}(x) \equiv \mathcal{L}_T - \mathcal{L}$, on can obtain (where $\rho, \lambda = 0,1,2,3$)

$$\delta\mathcal{L}(x) = \frac{i}{2} \varepsilon^{\rho\lambda} \bar\psi(x)[(\beta_\rho \partial_\lambda - \beta_\lambda \partial_\rho) - i[\beta^\mu, \Sigma_{\rho\lambda}]\partial_\mu]\psi(x). \qquad (a\text{-}4)$$

A necessary and sufficient condition that $\mathcal{L}$ be Lorentz invariant is $\delta\bar{\mathcal{L}}(x) = 0$, or equivalently,

$$\bar\psi(x)(\beta_\rho \partial_\lambda - \beta_\lambda \partial_\rho)\psi(x) = \bar\psi(x)(i[\beta^\mu, \Sigma_{\rho\lambda}]\partial_\mu)\psi(x). \qquad (a\text{-}5)$$

To prove $\delta\mathcal{L}(x) = 0$ or Eq. (a-5), we show that: (1) As $\rho = \lambda$, $\varepsilon^{\rho\lambda} = 0$, then $\delta\mathcal{L}(x) = 0$; (2) As $\rho = l$, $\lambda = m$ ($l, m = 1,2,3$), using Eqs. (3)-(4), $\Sigma_{lm} = \varepsilon_{lmn} I_{2\times 2} \otimes \tau_n$ as well as $[\tau_l, \tau_m] = i\varepsilon_{lmn} \tau_n$, one has $[\beta^0, \Sigma_{lm}] = 0$ and $i[\beta^\mu, \Sigma_{lm}]\partial_\mu = (\beta_l \partial_m - \beta_m \partial_l)$, thus Eq. (a-5) is true; (3) As $\rho = l = 1,2,3$, $\lambda = 0$ (or alternatively, $\rho = 0$, $\lambda = l = 1,2,3$), consider that $\Sigma_{l0} = -\Sigma_{0l} = i\chi_l$ and $i\beta^\mu \partial_\mu \psi(x) = 0$, Eq. (a-5) becomes

$$\bar{\psi}(x)[-(\boldsymbol{\beta} \cdot \partial)\chi_l + \beta^0 \partial^l]\psi(x) = 0.  \tag{a-6}$$

Using Eqs. (3)-(5) and the transversality condition $\partial^l E_l = \partial^l B_l = 0$, it is easy to show that Eq. (a-6) is true.

The statements (1), (2) and (3) exhaust all cases, therefore, the pseudo-Lagrangian density given by Eq. (17) is Lorentz invariant. In fact, one can also directly prove that Eq. (6) is Lorentz covariant by applying Eqs. (1) and (2).

## Appendix B

### Analogy between the free electron and the free electromagnetic field

In relativistic quantum mechanics, a free electron with mass $m$ is described by the Dirac equation for spin-1/2 particles ($\mu, \nu = 0,1,2,3$):

$$(i\gamma^\mu \partial_\mu - m)\varphi(x) = 0,  \tag{b-1}$$

where $\gamma^\mu$'s are the 4×4 Dirac matrices satisfying the algebra $\gamma^\mu \gamma^\nu + \gamma^\nu \gamma^\mu = 2g^{\mu\nu}$. Let the four-component spinor $\varphi$ be decomposed into two two-component spinors $\chi$ and $\phi$: $\varphi = \begin{pmatrix} \chi \\ \phi \end{pmatrix}$, in terms of the Pauli matrix vector $\boldsymbol{\sigma}$, the Eq.(b-1) can also be rewritten as the Maxwell-like form:

$$\begin{cases} (\boldsymbol{\sigma} \cdot \nabla)\chi = (-\partial_t + im)\phi \\ (\boldsymbol{\sigma} \cdot \nabla)\phi = (-\partial_t - im)\chi \end{cases}.  \tag{b-2}$$

Likewise, using Eqs. (3)-(5) the Dirac-like equation (6)

$$i\beta^\mu \partial_\mu \psi(x) = 0  \tag{b-3}$$

can also be rewritten as the matrix form of the Maxwell equations:

$$\begin{cases} (\boldsymbol{\tau} \cdot \nabla) B = i \partial_t E \\ (\boldsymbol{\tau} \cdot \nabla) E = -i \partial_t B \end{cases}. \qquad \text{(b-4)}$$

Therefore, the Maxwell equations for the free electromagnetic field can be rewritten as the Dirac-like equation; conversely, the Dirac equation for the free electron can be rewritten as the Maxwell-like equations. Comparing the Dirac equation Eq.(b-1) with the like-Dirac equation Eq.(b-3), or comparing the Maxwell-like equation Eq.(b-2) with the Maxwell equation Eq.(b-4), we can see that, $\chi$ is to $\phi$, as $E$ is to $B$. Furthermore, we have the following analogies:

1) In the Dirac field $\varphi = \begin{pmatrix} \chi \\ \phi \end{pmatrix}$, $\chi$ and $\phi$ are two different types of 2-component spinor, they respectively correspond to the inequivalent representations (1/2, 0) and (0, 1/2) of the Lorentz group, and the Dirac equation presents a relation between the two spinors. If we consider parity, then it is no longer sufficient to consider the 2-spinors $\chi$ and $\phi$ separately, but the 4-spinor $\varphi = \begin{pmatrix} \chi \\ \phi \end{pmatrix}$. This 4-spinor is an irreducible representation of the Lorentz group extended by parity (i.e., the $(1/2, 0) \oplus (0, 1/2)$ representation). Likewise, as for the photon field $\psi = \frac{1}{\sqrt{2}} \begin{pmatrix} E \\ iB \end{pmatrix}$, the column matrices $E$ and $iB$ are two different types of 3-component spinor, they respectively correspond to the inequivalent representations (1, 0) and (0, 1) of the Lorentz group, and the Dirac-like equation presents a relation between the two spinors. If we consider parity, then it is no longer sufficient to consider the 3-spinors $E$ and $iB$ separately, but the 6-spinor $\psi = \frac{1}{\sqrt{2}} \begin{pmatrix} E \\ iB \end{pmatrix}$. This 6-spinor is an irreducible representation of the Lorentz group extended by parity (i.e., the $(1, 0) \oplus (0, 1)$ representation).

2) By Eq.(b-4), a moving or time-varying electric field $E$ induces the magnetic field $B$ and vice versa. By Eq.(b-2), a moving or time-varying component $\phi$ induces the component $\chi$ and vice versa.

3) Under the exchange $E \leftrightarrow iB$, the free electromagnetic fields have the electricity-magnetism duality property. Under the exchange $\chi \leftrightarrow \phi$, the free Dirac fields have particle- antiparticle symmetry property.

4) The quantity $\boldsymbol{E}^2 - \boldsymbol{B}^2$ is Lorentz invariant; the quantity $\phi^+\phi - \chi^+\chi$ is also Lorentz invariant.

5) The quantity $\boldsymbol{E}^2 + \boldsymbol{B}^2$ is proportional to the density of photon number; the quantity $\phi^+\phi + \chi^+\chi$ is proportional to the density of electron number.

6) For the fields excited by electric source, the electric field $E$ is the large component and the magnetic field $B$ the small component, while for the fields excited by magnetic source, the magnetic field $B$ is the large component and the electric field $E$ the small component. Analogously, for the fields of negative electron, $\phi$ is the large component and $\chi$ the small component, while for the fields of positive electron, $\chi$ is the large component and $\phi$ the small component.

In a word, when a wave packet of electron is moving with high speeds or varies rapidly, or its size is sufficiently small, or in the presence of a strong electromagnetic field, its small components and the related effects cannot be ignored. Especially, a negative-electron wavepacket does contain a positive-electron component, and vice versa, just as that a moving or time-varying electric field is always accompanied by a magnetic field component, and vice versa.